\documentclass[
twocolumn,
preprintnumbers,amsmath,amssymb,prb
]{revtex4-1}

\usepackage{graphicx}
\usepackage{dcolumn}
\usepackage{bm}
\usepackage{amssymb}
\usepackage{epstopdf}
\usepackage{color}
\usepackage{amsmath}
\usepackage{ulem}
\DeclareMathOperator{\sign}{sign}

\begin{document}

\title{{Elementary excitations in the hybrid Bose-Fermi system induced by circularly polarized light in a two-dimensional gas of charge carriers with different masses}}

\author{V. M. Kovalev$^{1,2}$}\email{vadimkovalev(c)isp.nsc.ru}
\author{M. V. Boev$^{1}$}
\author{O. V. Kibis$^{1}$}

\affiliation{$^1$Department of Applied and Theoretical Physics,
Novosibirsk State Technical University, Karl Marx Avenue 20,
Novosibirsk 630073, Russia}
\affiliation{$^2$Abrikosov Center for Theoretical Physics, MIPT University, Dolgoprudny 141701, Russia}

\begin{abstract}
We developed a theory describing elementary excitations in the Bose-Fermi system induced by circularly polarized light in a two-dimensional (2D) gas of charge carriers with different masses. In such a hybrid system, the Fermi subsystem is a degenerate Fermi gas, whereas the Bose subsystem is a condensate of the light-induced composite bosons consisting of two fermions (electrons or holes) with different effective masses. The interaction of the single-particle excitations and the collective excitations (plasmons) in the Fermi subsystem with the Bogoliubov collective modes (bogolons) in the Bose subsystem is analyzed. The renormalization and damping (lifetime) of the excitations are calculated, and the possibility of their experimental observation is discussed. The developed theory can be applied to describe 2D condensed-matter structures containing charge carriers with different effective masses, including transition metal dichalcogenide monolayers and semiconductor quantum wells.
\end{abstract}

\maketitle

\section{Introduction}

All-optical control of electronic properties of condensed-matter structures by a high-frequency off-resonant electromagnetic field, which is based ideologically on the Floquet theory of periodically driven quantum systems (``Floquet engineering''), has become an established research area during last decades~\cite{Oka_2019,Basov_2017,Goldman_2014,Bukov_2015,Eckardt_2015,Casas_2001,Kibis_2020_1,Nuske_2020,Kibis_2022,Liu_2022,Seshadri_2022,Kobayashi_2023}.
Since the off-resonant field cannot be absorbed by electrons, it only dresses them, producing the composite electron-field states with unusual physical properties. Particularly, it has been demonstrated that such a dressing field can crucially modify electronic characteristics of various condensed-matter nanostructures,
including  semiconductor quantum wells~\cite{Lindner_2011}, quantum rings~\cite{Koshelev_2015}, quantum dots~\cite{Kryuchkyan_2017}, topological insulators~\cite{Rechtsman_2013,Wang_2013,Zhu_2023,Torres_2014}, carbon nanotubes~\cite{Kibis_2021_1},
graphene and related two-dimensional materials~\cite{Oka_2009,Syzranov_2013,Usaj_2014,Perez_2014,Sie_2015,Iurov_2019,Cavalleri_2020}, etc.

Among many phenomena induced by a dressing field, the Floquet engineering of electron behaviour in various potential reliefs takes deserved place. If the field is both strong and high-frequency, the electron dynamics can be described by the effective dressed potential which can be obtained from a ``bare'' potential by its averaging along the classical electron trajectory under the field over the field period. The most pronounced modification of the potentials takes place in low-dimensional electronic systems. Particularly,
the two-dimensional (2D) repulsive Coulomb potential under a circularly polarized dressing field acquires an attractive area in its core~\cite{Kibis_2019}, what leads to confinement of conduction electrons at repulsive potentials in quantum wells~\cite{Kibis_2020_2}. The same field-induced attraction can manifest itself in the processes of electron-electron interaction. Recently, it was demonstrated theoretically that the circularly polarized irradiation of two-dimensional conducting systems can produce composite bosons consisting of two electrons
with different effective masses~\cite{Kibis_2019}, which are stable due to the Fermi sea of conduction electrons~\cite{Kibis_2021_3}. As a result, an optically induced mixture of paired electrons and normal conduction electrons (the hybrid Bose-Fermi system) appears. Since the optically induced hybrid Bose-Fermi system~\cite{Kibis_2021_3} is interesting from viewpoint of possible light-induced superconductivity and superfluidity, the present article is aimed to study elementary excitations there.

Physical properties of nanostructures and their response to external perturbations are determined by the spectrum of elementary excitations. Evidently, the type of elementary excitations existing in various physical systems depends on the quantum statistics of initial bare particles filling the system. In the past, only two quantum systems of the Fermi-type were known: The electron gas in metals (or semiconductors) and liquid helium, $^3$He. As to the Bose-type liquid, its typical example was $^4$He. All these quantum objects have the rich spectra of elementary excitations determining their unique physical properties at low temperatures \cite{LL9, AGD, FW}. Other interesting quantum systems are presented by the mixtures of the Bose and Fermi gases. In  such mixtures, new interaction channels appear due to the interactions between bosons and fermions. As an example, a new type of paring between fermions due to the exchange by the excitations of the Bose subsystem may occur, including s-type \cite{bardeen, wang} and p-type \cite{kagan} Fermi-particles pairing. Historically, all these types of the hybrid Bose-Fermi systems were initially considered to be applied to cold atomic systems \cite{Ferrier, Ozawa,  Zheng1, Zhang2, Cui, Kinnunen, Tylutki}. However, the technological achievements in the design and fabrication of nanostructures have recently stimulated intensive theoretical discussions about new physical phenomena in the condensed-matter Bose-Fermi mixtures \cite{Cotlet, Kavokin, Kavokin2, Meng, Villegas,Boev_2020}. Particularly, a possibility of experimental realizations of long-living 2D dipolar exciton systems or 2D exciton-polariton gases opens a way to create the condensed matter Bose-Fermi mixtures, where the Bose subsystem is either an exciton or exciton-polariton gas. Thus, the physics of the hybrid Bose-Fermi systems in low-dimensional structures is the established research area of modern science, which forms the basis for the present study.

The specific renormalization of the Coulomb interaction between charged particles by a dressing field~\cite{Kibis_2019,Kibis_2021_3} opens a way to form the mixture of two subsystems, where the first one is the degenerate Fermi gas of light and heavy normal electrons, whereas the second one is the Bose gas consisting of the bound two-electron composite bosons. At low temperatures, the latter may form the Bose-Einstein condensate (BEC), where composite bosons interact via the short-range potential due to a strong screening of their direct Coulomb interaction by normal electrons. Thus, under the external irradiation by a circularly polarized electromagnetic field (see Fig.~1), light electrons attractively interact with heavy electrons to form the two-electron composite bosons being in the BEC regime, whereas the remaining unpaired electrons form the degenerate Fermi gas. {Certainly, it should be kept in mind that BEC in real systems depends on many additional physical factors (see, e.g., Ref.~\onlinecite{Wei_2023}) which should be analyzed carefully for samples planned to be studied experimentally.} In the present article, we consider the renormalization of physical properties of individual excitations in the Fermi subsystem of unpaired electrons and study the properties of various collective modes in the light-induced Bose-Fermi mixture, including the polaron effect, the quasi-particle lifetime, the renormalization of the collective mode dispersion laws and their damping.

The article is organized as follows. In Sec.~II, we describe the model under consideration and introduce the Hamiltonian describing the interaction between normal electrons (the Fermi subsystem) and the light-induced Bose subsystem consisting of paired electrons with different masses. In Sec.~III, the single-particle and collective modes in the optically induced hybrid Bose-Fermi systems are analyzed. The last two sections contain the conclusion and acknowledgements, whereas Appendix contains derivation of the interaction Hamiltonian for two electrons with different effective masses.
\begin{figure}[t]
\includegraphics[width=0.8\columnwidth]{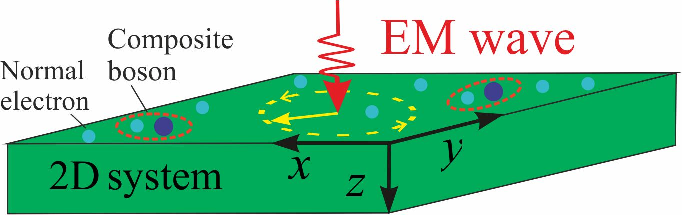}
\caption{Sketch of the system under consideration: A two-dimensional system containing heavy electrons (large circles) and light electrons (small circles) under irradiation by a circularly polarized electromagnetic wave. As a result of the irradiation, the hybrid Bose-Fermi system consisting of composite bosons (paired heavy and light electrons) and the degenerate Fermi gas of normal electrons appears.}
\label{Fig.1}
\end{figure}
\begin{figure}[t]
\includegraphics[width=0.8\columnwidth]{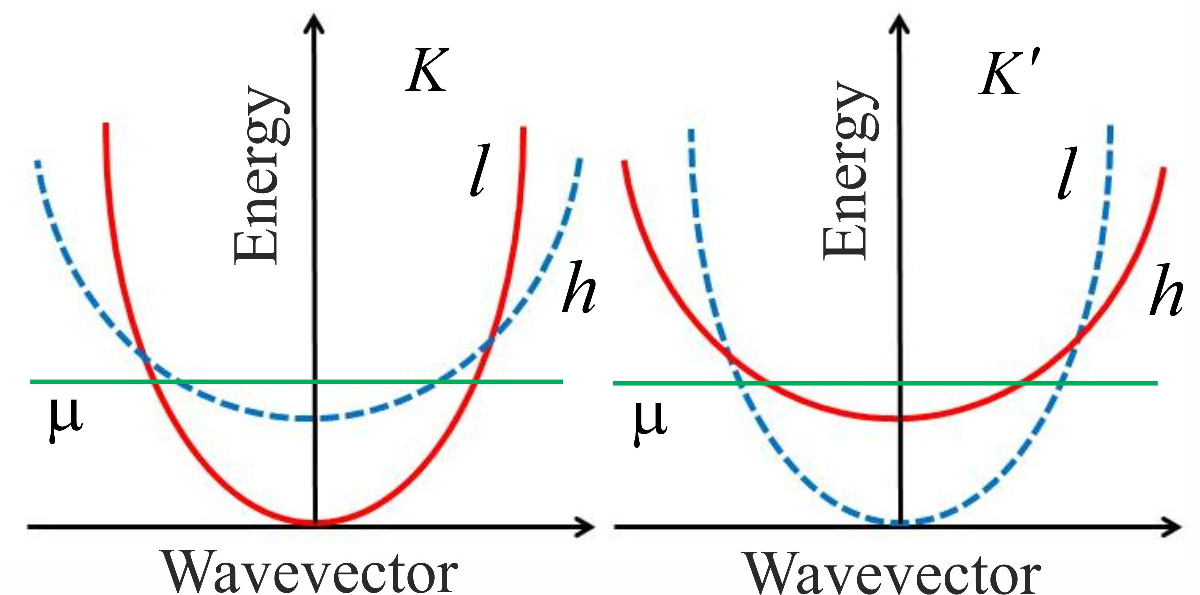}
\caption{Structure of the conduction band in MoS$_2$ monolayer:   The electron energy spectrum in the two valleys ($K$ and $K'$) consists of the branches of heavy ({\it h}) and light ({\it l}) electrons with the mutually opposite spin orientation (the solid and dashed lines), where $\mu$ is the Fermi energy.}
\label{Fig.1.1}
\end{figure}

\section{Model}
As it has been noted above, the light-induced Bose-Fermi system may occur in nanostructures containing charge carriers with different effective masses. For definiteness, we consider the electronic system in such a transition metal dichalcogenide material as MoS$_2$ monolayer which is under active study nowadays, showing unique optical and transport properties~\cite{Falko}. The conduction band of this material consists of the two non-equivalent valleys in the $K$ and $K'$ points of the Brillouin zone, where each valley contains the two spin-split electron branches corresponding to the heavy electrons with the mass $m_h=0.46m_0$ and the light electrons with the mass $m_l=0.43m_0$, where $m_0$ is the free electron mass (see Fig.~2). As a consequence, the circularly polarized irradiation of the monolayer may form the Bose subsystem consisting of two electrons with different effective masses~\cite{Kibis_2019,Kibis_2021_3}, which can be considered as composite bosons with the effective mass $M=m_h+m_l$ and the charge $2e$, where $e=-|e|$ is the electron charge. In the following, we will assume the boson density to be small enough to consider the Bose subsystem as a gas of weakly interacting composite bosons. For the Fermi level plotted in Fig.~2, the Fermi subsystem contains unpaired light and heavy electrons. However, the density of light electrons much exceeds the density of heavy electrons since the ground branch corresponds to light electrons. To simplify the consideration of the light-induced Bose-Fermi mixture, we will neglect the contribution of heavy electrons into the Fermi subsystem and will assume that the Fermi subsystem consists only of light electrons with the mass $m=m_l$. Another simplification of the model is related to the two valley structure of the MoS$_2$ Brillouin zone. Namely, we will not take into account the intervalley scattering processes because they require the extremely large momenta transfer between interacting particles, whereas all phenomena considered below occur in the long-wavelength limit corresponding to very small momenta.

{Since the Hamiltonian describing the light-induced electron pairing was analyzed earlier~\cite{Kibis_2019,Kibis_2021_3} (see Appendix for details), the following analysis is devoted to the Hamiltonian describing the interaction processes in the light-induced Bose-Fermi system. Conventionally, the interactions of charge particles in various 2D structures are described by the two-dimensional Coulomb potential (see, e.g., Refs.~\onlinecite{Lifshitz_book,Kotov_2012}). Therefore, the interaction Hamiltonian for the considered hybrid Bose-Fermi system can be written as a sum of three terms, $H=H_{BF}+H_{FF}+H_{BB}$, where the term
\begin{eqnarray}\label{Eq1}
H_{BF}=\int_S d^2\textbf{r}\int_S d^2\textbf{R}\,\hat{n}(\textbf{r},t)U_{BF}(\textbf{r}-\textbf{R})\hat{N}(\textbf{R},t),
\end{eqnarray}
describes the interaction between the Fermi and Bose subsystems with the two-dimensional Coulomb potential
\begin{equation}\label{BF}
U_{BF}(\textbf{r}-\textbf{R})=\frac{2e^2}{\epsilon|\textbf{r}-\textbf{R}|},
\end{equation}
the term
\begin{equation}\label{Eq2}
H_{FF}=\frac{1}{2}\int_S d^2\textbf{r}\int_S d^2\textbf{r}'\,\hat{n}(\textbf{r},t)U_{FF}(\textbf{r}-\textbf{r}')\hat{n}(\textbf{r}',t),
\end{equation}
describes the interactions of fermions within the Fermi subsystem with the two-dimensional Coulomb potential
\begin{equation}\label{FF}
U_{FF}(\textbf{r}-\textbf{r}')=\frac{e^2}{\epsilon|\textbf{r}-\textbf{r}'|},
\end{equation}
the term
\begin{equation}\label{Eq22}
H_{BB}=\frac{1}{2}\int_S d^2\textbf{R}\int_S d^2\textbf{R}'\,\hat{N}(\textbf{R},t)U_{BB}(\textbf{R}-\textbf{R}')\hat{N}(\textbf{R}',t),
\end{equation}
describes the interactions of composite bosons within the Bose subsystem screened by normal electrons with the two-dimensional screened Coulomb potential
\begin{equation}\label{BB}
U_{BB}(\textbf{R}-\textbf{R}')=\int \frac{d^2\mathbf{k}}{(2\pi)^2}\,U_B(\mathbf{k})e^{i\mathbf{k}(\mathbf{R}-\mathbf{R}')},
\end{equation}
the Fourier image of the screened potential is
\begin{equation}\label{UB}
U_B(\mathbf{k})=\frac{8\pi e^2}{\epsilon(k+k_s)}
\end{equation}
$k_s=2/a_s$ is the Thomas-Fermi screening wavenumber, $a_s=\epsilon\hbar^2/me^2$ is the effective screening length, $\epsilon$ is the effective dielectric constant accounting for all effects of medium,
$\hat{n}(\textbf{r},t)=\psi^\dag(\textbf{r},t)\psi(\textbf{r},t)$ and $\hat{N}(\textbf{R},t)=\varphi^\dag(\textbf{R},t)\varphi(\textbf{R},t)$ are the density operators of the Fermi and Bose subsystem, respectively, $\textbf{R}=(x,y)$ is the plane radius vector of {composite boson}, $\textbf{r}=(x,y)$ is the plain radius vector of normal electron, and $S$ is the area of the 2D system.}

We will restrict the following consideration by the case of extremely low temperatures, assuming that {composite bosons} form BEC, whereas the remaining unpaired electrons form the normal degenerate Fermi gas. In the following, we will also assume that the BEC density $n_c$ is small enough to satisfy the condition $n_cl^2\ll1$, where $l$ is the boson-boson scattering length. In such a regime, BEC can be described by the standard Bogoliubov theory of weakly interacting Bose gas.
Within this theory, the Bose operator $\varphi(\textbf{R},t)=\varphi_0+\delta\varphi(\textbf{R},t)$ consists of the uniform part describing BEC and the fluctuating part, where $|\varphi_0|^2=n_c$ is the BEC density. As a result, the interaction Hamiltonian \eqref{Eq1} can be rewritten as a sum of the three terms,
\begin{align}\label{Eq33}
&H^{(0)}_{BF}=n_c\int_S d^2\mathbf{r}\int_S d^2\textbf{R}\hat{n}(\textbf{r},t)U_{BF}(\textbf{r}-\textbf{R}),\nonumber\\
&H^{(1)}_{BF}=\sqrt{n_c}\int_S d^2\mathbf{r}\int_S d^2\textbf{R}\hat{n}(\textbf{r},t)U_{BF}(\textbf{r}-\textbf{R})\nonumber\\
&\times\left[\delta\varphi^*(\textbf{R},t)+\delta\varphi(\textbf{R},t)\right],\nonumber\\
&H^{(2)}_{BF}=\int_S d^2\mathbf{r}\int_S d^2\textbf{R}\hat{n}(\textbf{r},t)U_{BF}(\textbf{r}-\textbf{R})|\delta\varphi(\textbf{R},t)|^2.
\end{align}
where the first term, which describes the shift of the Fermi energy of unpaired electrons, does not affect electronic properties and will be omitted in the following, whereas the second and third terms describe the interaction of the Fermi subsystem with the Bogoliubov excitations of BEC (bogolons). For further developments, it is instructive to introduce the creation and annihilation operators for bogolons~\cite{LL9} via the relations
\begin{eqnarray}\label{Eq4}
\delta\varphi(\textbf{R},t)=\frac{1}{\sqrt{S}}\sum_{\bf p}e^{i{\bf pR}}\left(u_{\bf p}b_{\bf p}+v_{\bf p}b^\dag_{-{\bf p}}\right),\nonumber\\
\delta\varphi^*(\textbf{R},t)=\frac{1}{\sqrt{S}}\sum_{\bf p}e^{-i{\bf pR}}\left(u^*_{\bf p}b^\dag_{\bf p}+v^*_{\bf p}b_{-{\bf p}}\right),
\end{eqnarray}
where $u_{\bf p},v_{\bf p}$ are the standard Bogoliubov coefficients (here and below we use the system of units with $\hbar=1$ and will restore the Plank constant in the final expressions only). Introducing the healing length $\zeta=1/2Ms$, the Bogoliubov coefficients read
\begin{gather}\label{Eq4.1}
u_{{\bf p}}, v_{{\bf p}}=\pm\sqrt{\frac{p^2/2M+U_B n_c}{2\omega_{{\bf p}}}\pm\frac{1}{2}},\\\nonumber
\omega_{{\bf p}}=sp\sqrt{1+(p\zeta)^2},
\end{gather}
where $\omega_{{\bf p}}$ is the bogolon dispersion, the parameter $U_B\equiv U_B(\textbf{k}=0)$ represents the strength of the boson-boson interaction, and  $s=\sqrt{U_B n_c/M}$ is the bogolon phase velocity. With using the bogolon operators, the interaction terms~\eqref{Eq33} can be rewritten as
\begin{align}\label{Eq4.2}
&H^{(1)}_{BF}=\sqrt{\frac{n_c}{S}}\sum_{{\bf p}}U^{BF}_{\mathbf{p}}\hat{n}_{-\mathbf{p}}\nonumber\\
&\times\left[(u_{\bf p}+v_{-{\bf p}})b_{\bf p}+(u_{-{\bf p}}+v_{\bf p})b^\dag_{-{\bf p}}\right],
\end{align}
and
\begin{align}\label{Eq4.3}
&H^{(2)}_{BF}=\frac{1}{S}\sum_{{\bf k},{\bf p}}U^{BF}_{\mathbf{p}}\hat{n}_{-\mathbf{p}}\nonumber\\
&\times\left(u_{{\bf k}-{\bf p}}b^\dag_{{\bf k}-{\bf p}}+v_{{\bf k}-{\bf p}}b_{{\bf p}-{\bf k}}\right)\left(u_{\bf k}b_{\bf k}+v_{\bf k}b^\dag_{-{\bf k}}\right),
\end{align}
where the term \eqref{Eq4.2} describes the fermion-boson interaction with a single bogolon, whereas the term \eqref{Eq4.3} corresponds to the two-bogolon processes. Mathematically, the interaction $H^{(1)}_{BF}$ is similar to the conventional electron-phonon interaction in normal electronic systems (the only difference is the Bogoliubov coefficients), whereas the second term $H^{(2)}_{BF}$ essentially differs from the usual electron-phonon Hamiltonian. Nevertheless, the terms \eqref{Eq4.2} and \eqref{Eq4.3} are of the same order and should be considered simultaneously.
\begin{figure}[t]
\includegraphics[width=1.0\columnwidth]{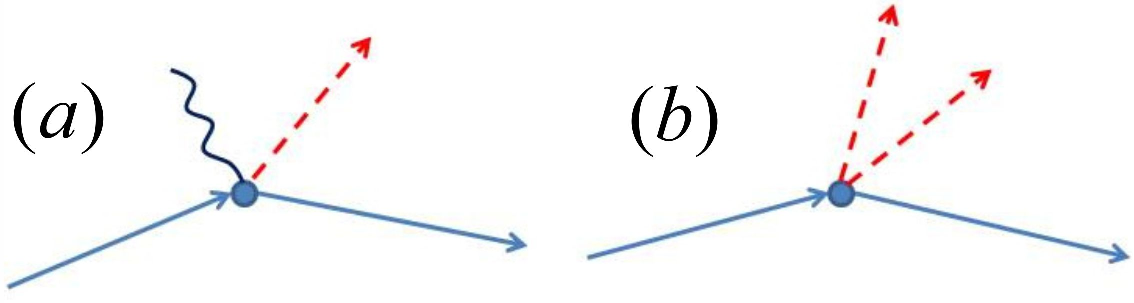}
\caption{Vertex Feynman diagrams describing the amplitudes of the single bogolon (a) and the double bogolon (b) emission by a moving fermion. The solid lines correspond to fermions, the dashed lines correspond to the Bogoliubov excitations, the wavy lines mark the factor $\sqrt{n_c}$, and the circles mark the boson-fermion interaction potential.}
\label{Fig.2}
\end{figure}

The Feynman diagrams corresponding to the quantum amplitudes of the processes described by the Hamiltonians (\ref{Eq4.2}) and (\ref{Eq4.3}) are presented in Fig.~3. It should be noted that these processes can be considered separately in the lowest order with respect to the boson-fermion interaction potential $U^{BF}_{\mathbf{p}}$. Then the correction to the bare fermion energy,
\begin{equation}\label{xi}
\xi_{\mathbf{p}}=p^2/2m-\mu,
\end{equation}
which appears due to the interactions (\ref{Eq4.2}) and (\ref{Eq4.3}), is given by the self-energy contribution $\Sigma(\varepsilon,\textbf{p})$ to the pole of the fermion Green function
$\mathcal{G}^{-1}(\varepsilon,\mathbf{p})=\varepsilon-\xi_\mathbf{p}+i\delta \sign(\xi_\mathbf{p})$. As to the renormalized fermion dispersion, it is determined by the equation $\varepsilon-\xi_\mathbf{p}-\Sigma(\varepsilon,\textbf{p})=0$.  The self-energy, $\Sigma(\varepsilon,\textbf{p})=\Sigma_{cn}(\varepsilon,\textbf{p})+\Sigma_{nn}(\varepsilon,\textbf{p})$, makes the two contributions in the second order of the fermion-boson interaction potential $U^{BF}_\mathbf{k}$, which are
\begin{align}\label{Eq4.4}
&\Sigma_{cn}(\varepsilon,\textbf{p})=i\sum_{\omega,\mathbf{k}}|U^{BF}_k|^2
\mathcal{G}(\varepsilon-\omega,\textbf{p}-\mathbf{k})P_{cn}(\omega,\textbf{k}),\nonumber\\
&P_{cn}(\omega,\textbf{k})=n_c\Bigl[G(\omega,\textbf{k})+\tilde{G}(\omega,\textbf{k})+
F(\omega,\textbf{k})+\tilde{F}(\omega,\textbf{k})\Bigr],
\end{align}
and
\begin{align}\label{Eq4.5}
&\Sigma_{nn}(\varepsilon,\textbf{p})=i\sum_{\omega,\mathbf{k}}|U^{BF}_k|^2\mathcal{G}(\varepsilon-\omega,\textbf{p}-\mathbf{k})P_{nn}(\omega,\textbf{k}),\nonumber\\
&P_{nn}(\varepsilon,\textbf{p})=i\sum_{\omega,\mathbf{k}}\Bigl[G(\varepsilon,\mathbf{p})G(\varepsilon-\omega,\mathbf{p}-\mathbf{k})\nonumber\\
&+\tilde{G}(\varepsilon,\mathbf{p})\tilde{G}(\varepsilon-\omega,\mathbf{p}-\mathbf{k})\nonumber\\
&+F(\varepsilon,\mathbf{p})\tilde{F}(\varepsilon-\omega,\mathbf{p}-\mathbf{k})+
\tilde{F}(\varepsilon,\mathbf{p})F(\varepsilon-\omega,\mathbf{p}-\mathbf{k})\Bigr],
\end{align}
where the Green functions of BEC read
\begin{eqnarray}\label{Eq4.6}
G(\varepsilon,\mathbf{p})&=&\frac{\varepsilon+p^2/2M+U_{B}n_c}{\varepsilon^2-\omega^2_\mathbf{p}+i\delta},\nonumber\\
F(\varepsilon,\mathbf{p})&=&\frac{-U_{B}n_c}{\varepsilon^2-\omega^2_\mathbf{p}+i\delta},
\end{eqnarray}
and $\tilde{G}(\varepsilon,\mathbf{p})=G(-\varepsilon,-\mathbf{p}),\,\,\,\tilde{F}(\varepsilon,\mathbf{p})=F(-\varepsilon,-\mathbf{p})$.
Physically, the self-energy $\Sigma_{cn}(\varepsilon,\textbf{p})$ describes the excitation of BEC accompanied by transition of a boson to the non-condensed state (see Fig.~4a) and arises from the interaction Hamiltonian \eqref{Eq4.2}, whereas the self-energy $\Sigma_{nn}(\varepsilon,\textbf{p})$ describes the polarization of non-condensed composite bosons (see Fig.~4b) arisen from the interaction Hamiltonian \eqref{Eq4.3}.

{The self-energy operators (\ref{Eq4.4}) and (\ref{Eq4.5}) read
\begin{gather}\label{AppA1}
\Sigma_{cn(nn)}(\varepsilon,\textbf{p})=i\sum_{\omega,\mathbf{k}}\mathcal{G}(\varepsilon-\omega,\textbf{p}-\mathbf{k})R(\omega,\textbf{k}),
\end{gather}
where $R(\omega,\textbf{k})$ is either $|U^{BF}_k|^2P_{cn}(\omega,\textbf{k})$ or $|U^{BF}_k|^2P_{nn}(\omega,\textbf{k})$. In both cases, $R(\omega,\textbf{k})$ is the even function of frequency $\omega$ and depends on the absolute value of momentum $\textbf{k}$. Using this, Eq.~\eqref{AppA1} can be simplified. Namely, using the expression
\begin{align}\label{AppA2}
&\int\limits_0^{2\pi}\frac{d\varphi}{2\pi}\frac{1}{a+b\cos\varphi\pm i\delta}\nonumber\\
&=\frac{\sign(a)\theta[|a|-|b|]}{\sqrt{a^2-b^2}}\mp i\frac{\theta[|b|-|a|]}{\sqrt{b^2-a^2}},
\end{align}
we arrive at
\begin{align}\nonumber
&\Sigma(\varepsilon,\textbf{p})=i\int\limits_{-\infty}^{\infty}\frac{d\omega}{2\pi}\int\limits_0^{\infty}\frac{kdk}{2\pi}R(\omega,\textbf{k})
\Bigl[A(\omega,k)-iB(\omega,k)\Bigr],\\
&A(\omega,k)=\frac{\sign[\varepsilon+\omega-\xi_p-k^2/2m]}{\sqrt{(\varepsilon+\omega-\xi_p-k^2/2m)^2-v^2k^2}},\nonumber\\
&B(\omega,k)=\frac{\sign[\varepsilon+\omega]}{\sqrt{v^2k^2-(\varepsilon+\omega-\xi_p-k^2/2m)^2}}.
\label{AppA3}
\end{align}
In a vicinity of the Fermi level ($p\approx p_F$) and on the mass shell ($\varepsilon=\xi_p$), the small contribution of $k^2/2m$ can be ignored. Then Eqs.~\eqref{AppA3} can be written as
\begin{gather}
A(\omega,k)=\frac{\sign[\omega]}{\sqrt{\omega^2-v_F^2k^2}},\nonumber\\
B(\omega,k)=\frac{\sign[\varepsilon+\omega]}{\sqrt{v_F^2k^2-\omega^2}}.
\label{AppA4}
\end{gather}
Correspondingly, Eq.~\eqref{AppA1} yields
\begin{gather}\label{AppA5}
\Sigma(\varepsilon,\textbf{p})=\frac{\sign(\varepsilon)}{2\pi^2}\int\limits_0^{|\varepsilon|}d\omega\int\limits_{\omega/v_F}^{\infty}
\frac{kdk}{\sqrt{v_F^2k^2-\omega^2}}R(\omega,\mathbf{k}),
\end{gather}
whereas Eqs.~\eqref{Eq4.4} and \eqref{Eq4.5} read
}
\begin{align}\label{Eq4.7}
&\Sigma_{cn(nn)}(\varepsilon,\textbf{p})=\frac{\sign(\varepsilon)}{2\pi^2}\nonumber\\
&\times\int\limits_0^{|\varepsilon|}d\omega\int\limits_{\omega/v_F}^{\infty}
\frac{kdk|U_k^{BF}|^2}{\sqrt{v_F^2k^2-\omega^2}}P_{cn(nn)}(\omega,\mathbf{k}).
\end{align}
The most interesting case corresponds to the long-wavelength limit ($k\zeta\ll1$), when the Bogoliubov excitations have the linear sound-like dispersion, $\omega_k=sk$. Applying the Debye approximation, we will assume the linear dispersion $\omega_k=sk$ for all wavevectors $k$. This simplification has a great advantage enabling the analytical treatment of the problems under consideration below.

\begin{figure}[t]
\includegraphics[width=1.0\columnwidth]{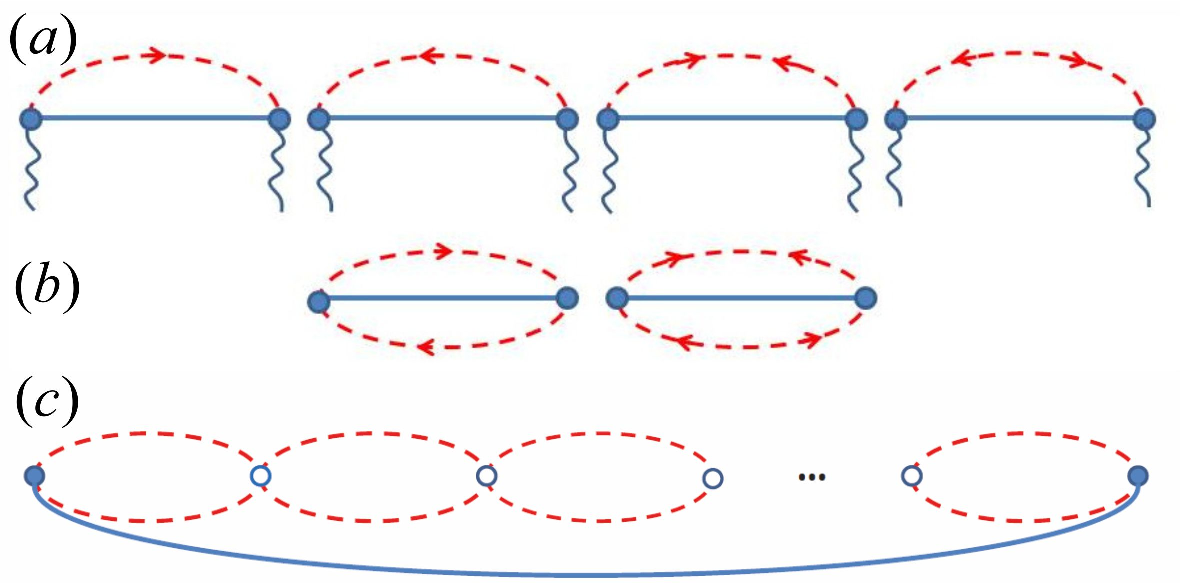}
\caption{Self-energy diagrams: (a) excitation of a boson to the non-condensed state by a moving fermion; (b) polarization of non-condensed bosons by a moving fermion; (c) infinite series of bubble diagrams contribution to $\textmd{Re}\,\Sigma_{nn}$. The solid lines corresponds to the electron Green functions, the dashed lines correspond to the bogolon Green functions, the wavy lines represent the $\sqrt{n_c}$ factor, the filled circles mark the boson-fermion interaction potential, and the empty circles mark the boson-boson interaction potential.}
\label{Fig.3}
\end{figure}

\section{Results and discussion}

\subsection{Polaron effect}

The polaron effect consists in the renormalization of the fermion effective mass in a vicinity of the Fermi energy due to the fermion-boson interaction and is described by the real part of the self-energy, $\textmd{Re}\,\Sigma(\varepsilon,\textbf{p})$. The solution of the equation $\varepsilon-\xi_\mathbf{p}-\textmd{Re}\,\Sigma(\varepsilon,\textbf{p})=0$ can be found in a vicinity of the Fermi energy by the successive approximation $\varepsilon=\xi_\mathbf{p}+\textmd{Re}\,\Sigma(\xi_\mathbf{p},p_F)$, where $p_F=\sqrt{2m\mu}$ is the Fermi momentum. Assuming the polaron corrections to be small, the terms $\textmd{Re}\,\Sigma_{cn}(\xi_\mathbf{p},p_F)$ and $\textmd{Re}\,\Sigma_{nn}(\xi_\mathbf{p},p_F)$ can be treated independently as follows.

The first term reads
\begin{align}
&\textmd{Re}\,\Sigma_{cn}(\xi_{\mathbf{p}},p_F)=\frac{\sign(\xi_{\mathbf{p}})}{2\pi^2}\nonumber\\
&\times\int\limits_0^{|\xi_{\mathbf{p}}|}d\omega\int\limits_{\omega/v_F}^{\infty}
\frac{kdk|U_k^{BF}|^2}{\sqrt{v_F^2k^2-\omega^2}}\textmd{Re}\,P_{cn}(\omega,\mathbf{k}),\nonumber\\
&\textmd{Re}\,P_{cn}(\omega,\mathbf{k})=n_c\frac{k^2}{M}\frac{1}{\omega^2-\omega_k^2},
\label{Polaron1}
\end{align}
where $v_F=p_F/m$ is the Fermi velocity. The integrals in Eq.~\eqref{Polaron1} can be easily evaluated in a vicinity of the Fermi energy, where $\xi_{\mathbf{p}}\rightarrow 0$. Substituting $\omega=0$ into the integral and taking into account
the screened boson-fermion interaction potential $U_k^{BF}=4\pi e^2/\epsilon(k+k_s)$,
one can find $\textmd{Re}\,\Sigma_{cn}(\xi_{\mathbf{p}},{p_F})=-b_{cn}\xi_{\mathbf{p}}$, where $b_{cn}$ is described by the expression
\begin{align}
&b_{cn}=\frac{n_c}{2\pi^2Ms^2v_F}\int\limits_{0}^{\infty}dk|U_k^{BF}|^2
=\frac{n_c}{2\pi^2Ms^2}\frac{(4\pi e^2)^2}{\epsilon^2\hbar v_Fk_s}\nonumber\\
&=\frac{e^2}{\epsilon\pi\hbar v_F}
\label{Polaron2}
\end{align}
with the restored Planck constant. The second correction to the fermion effective mass comes from the remaining self-energy part which reads
\begin{align}\label{Polaron3}
&\textmd{Re}\,\Sigma_{nn}(\xi_{\mathbf{p}},p_F)=\frac{\sign(\xi_{\mathbf{p}})}{2\pi^2}\nonumber\\
&\times\int\limits_0^{|\xi_{\mathbf{p}}|}d\omega\int\limits_{\omega/v_F}^{\infty}
\frac{kdk|U_k^{BF}|^2}{\sqrt{v_F^2k^2-\omega^2}}\textmd{Re}\,P_{nn}(\omega,\mathbf{k}),
\end{align}
where the polarization operator for non-condensed bosons is
\begin{gather}
P_{nn}=-\frac{(Ms)^2}{4}\left[\frac{1}{\sqrt{s^2k^2-\omega^2}}+i\frac{1}{\sqrt{\omega^2-s^2k^2}}\right].
\label{Polaron4}
\end{gather}
Substituting the real part of Eq.~\eqref{Polaron4} into Eq.~\eqref{Polaron3}, one can demonstrate that
$\textmd{Re}\,\Sigma_{nn}(\xi_{\mathbf{p}},p_F)\propto\xi_{\mathbf{p}}\ln\xi_{\mathbf{p}}$. Such a logarithmic divergence at $\xi_{\mathbf{p}}\rightarrow0$ means that the bubble diagrams pictured in Fig.~4b give a large contribution in a vicinity of Fermi energy. Therefore,  correct description of the interaction requires the summation of the infinite series of bubble diagrams pictured in Fig.~4c.
As a result of the summation, we arrive at the expression
\begin{align}
\label{Polaron5}
&\textmd{Re}\,\Sigma_{nn}(\xi_{\mathbf{p}},p_F)=\frac{\sign(\xi_{\mathbf{p}})}{2\pi^2v_F}\nonumber\\
&\times\int\limits_0^{|\xi_{\mathbf{p}}|}d\omega\int\limits_{0}^{\infty}
dk|U_k^{BF}|^2\frac{\textmd{Re}\,P_{nn}(0,\mathbf{k})}{1-U_k^{B}\textmd{Re}\,P_{nn}(0,\mathbf{k})},
\end{align}
which again has the form $\textmd{Re}\,\Sigma_{nn}(\xi_{\mathbf{p}},p_F)=-b_{nn}\xi_{\mathbf{p}}$ with $b_{nn}=b_{cn}{\cal F}(k_0/k_s)$, where $k_0^2=2\pi e^2(Ms)^2/\epsilon s\hbar^3\propto\sqrt{n_c}$ and the function
%
%
%
%
%
%
%
%
%
\begin{gather}
\label{Polaron7}
\mathcal{F}(y)=y^2\int\limits_0^\infty\frac{dx}{(x+1)^2(x+y^2)},
\end{gather}
describes the relationship between the BEC density, $n_c$, and the ratio $b_{nn}/b_{cn}$ (see Fig.~5). Since {the fermion energy reads $\varepsilon=(1-b_{cn}-b_{nn})\xi_{\mathbf{p}}$, the renormalized effective mass of fermion is
\begin{equation}\label{mm}
m^{*}=\frac{m}{1-b_{cn}-b_{nn}}\approx m(1+b_{cn}+b_{nn}),
\end{equation}
where the coefficients $b_{cn}$ and $b_{nn}$ are defined by Eq.~\eqref{Polaron2} and Fig.~6. It follows from Eq.~\eqref{mm} that the fermion-boson interaction leads to increasing the fermion effective mass, $m^{*}>m$. Particularly, $m^{*}\approx1.52m$ for the MoS$_2$ monolayer with the fermion and boson densities, $n=5\cdot10^{12}$~cm$^{-2}$ and $n_c=10^{8}$~cm$^{-2}$, respectively. The polaron renormalization of the effective mass will lead to decreasing electron mobility, what can manifest itself in various transport phenomena.

\begin{figure}[t]
\includegraphics[width=0.8\columnwidth]{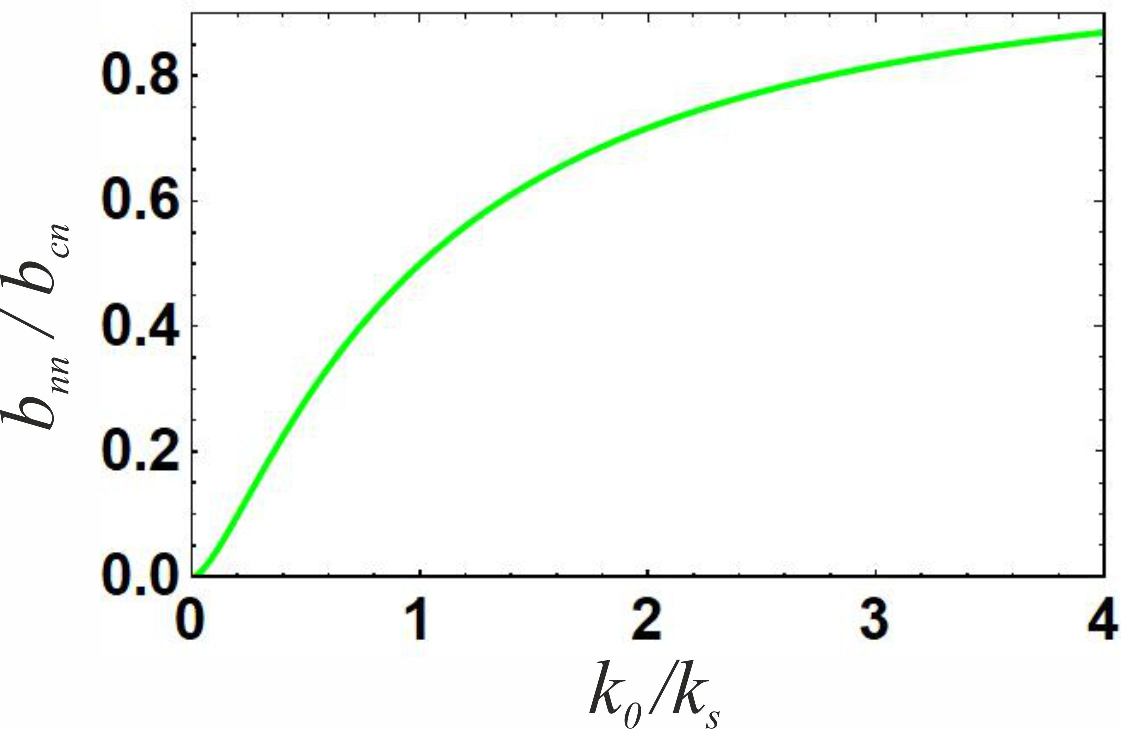}
\caption{Dependence of the ratio $b_{nn}/b_{cn}$ on the ratio $k_0/k_s$.}
\label{Fig.5}
\end{figure}
{It should be noted that the small parameter of the renormalization theory developed above is the ratio $e^2/\epsilon\hbar v_F$ , where the Fermi velocity $v_F$  can be increased by the gate voltage applied to a monolayer up to the electron density $\sim10^{14}$~cm$^{-2}$ (see. e.g., Ref.~\onlinecite{Wakarsuki_2017}), and the effective dielectric constant $\epsilon$ can be increased if the monolayer is sandwiched by dielectric materials with large dielectric constants. As a consequence, the aforesaid parameter can be varied in broad range to keep the obtained results within applicability of the renormalization theory. However, even if the calculated fermion mass lies near the border of applicability of the renormalization theory, the obtained results stay to be useful, at least, for semi-qualitative estimations.}

\subsection{Quasi-particle lifetime}

The imaginary part of fermion self-energy, $\textmd{Im}\,\Sigma(\xi_{\mathbf{p}},p_F)=-\Gamma$, defines the quasi-particle damping rate $\Gamma=1/2\tau_e$, where $\tau_e$ is the quasi-particle lifetime. In the following, we will analyse the damping rate,
\begin{equation}\label{GG}
\Gamma=\Gamma_{cn}+\Gamma_{nn},
\end{equation}
coming from the two contributions to the self-energy. The first contribution,
\begin{align}\nonumber
&\textmd{Im}\,\Sigma_{cn}(\xi_{\mathbf{p}},p_F)=\frac{\sign(\xi_{\mathbf{p}})}{2\pi^2}\\
&\times\int\limits_0^{|\xi_{\mathbf{p}}|}d\omega\int\limits_{\omega/v_F}^{\infty}
\frac{kdk|U_k^{BF}|^2}{\sqrt{v_F^2k^2-\omega^2}}\textmd{Im}\,P_{cn}(\omega,\mathbf{k}),\\
&\textmd{Im}\,P_{cn}(\omega,\mathbf{k})=-\pi n_c\frac{k^2}{M}\delta(\omega^2-\omega_k^2),
\label{lifetime1}
\end{align}
yields
\begin{align}\label{lifetime2}
&\Gamma_{cn}=\frac{\sign(\xi_{\mathbf{p}})n_c(4\pi e^2)^2\theta[v_F-s]}{4\pi\epsilon^2\hbar^2 Ms\sqrt{v_F^2-s^2}}\nonumber\\
&\times\left[\ln\left(\frac{|\xi_{\mathbf{p}}|+\hbar sk_s}{\hbar sk_s}\right)-\frac{|\xi_{\mathbf{p}}|}{|\xi_{\mathbf{p}}|+\hbar sk_s}\right],
\end{align}
where $\theta[x]$ is the Heaviside step-function (the Planck constant is restored). The second contribution has the form
\begin{align}\label{lifetime3}
&\textmd{Im}\,\Sigma_{nn}(\xi_{\mathbf{p}},p_F)=-\frac{\sign(\xi_{\mathbf{p}})(Ms)^2}{8\pi^2}\nonumber\\
&\times\int\limits_0^{|\xi_{\mathbf{p}}|}d\omega\int\limits_{\omega/v_F}^{\omega/s}
\frac{kdk|U_k^{BF}|^2}{\sqrt{v_F^2k^2-\omega^2}\sqrt{\omega^2-s^2k^2}}.
\end{align}
Since $|U_k^{BF}|^2\approx|U_0^{BF}|^2$ for $|\xi_{\mathbf{p}}|\ll {v_F}k_s$, Eq.~\eqref{lifetime3} yields
\begin{gather}\label{lifetime4}
\Gamma_{nn}=\xi_{\mathbf{p}} \frac{(Ms)^2}{16\pi v_Fs\hbar^4}\left(\frac{4\pi e^2}{\epsilon k_s}\right)^2,
\end{gather}
where the Planck constant is restored. One can see that both contributions are non-zero only if the fermion phase velocity $v_F$ exceeds the velocity of Bogoliubov excitations, $s$. Physically, this is the condition of bogolon emission by a fermion (the particular case of the Cherenkov effect). Thus, the damping arises from the emission of bogolons which are real (in contrast to the polaron effect discussed above, where the Bogoliubov excitations dressing a fermion are virtual).

{In the aforesaid, we took into account only the bogolon damping arisen for the bogolon-fermion interaction, although there is also the bogolon-bogolon interaction channel giving the additional contribution to the decay rate known as the Beliaev dumping (see, e.g., Ref.~\onlinecite{Chung_2009}). However, the Beliaev dumping is $\sim k^3$ in the long-wavelength limit considered above, whereas the dumping arisen from the bogolon-fermion interaction is $\sim k$ there. Therefore, the Beliaev dumping can be neglected as a first approximation.}

It should be noted that the quasi-particle description holds only if the damping is weak enough, $\Gamma/\xi_{\mathbf{p}}\ll1$. To validate this condition, it should be noted that $|\xi_{\mathbf{p}}|\ll {v_F}k_s$ in a vicinity of Fermi energy. As a consequence, $\Gamma_{cn}\propto \xi_{\mathbf{p}}|\xi_{\mathbf{p}}|$  and, therefore, $\Gamma_{cn}/\xi_{\mathbf{p}}\ll1$.  Thus, the processes corresponding to the boson transfer from the condensate into non-condensed states due to the moving fermion (see Fig.~\ref{Fig.2}a and Fig.~\ref{Fig.3}a) do not destroy the quasi-particle description of the Fermi subsystem. Substituting the MoS$_2$ monolayer parameters~\cite{Falko} into Eq.~\eqref{lifetime4}, one can see that  $\Gamma_{nn}/\xi_{\mathbf{p}}\sim0.04$ and, therefore, the condition $\Gamma_{nn}/\xi_{\mathbf{p}}\ll1$ is also satisfied.

It should be noted also that the electron gas viscosity is directly related to the electron-electron scattering time~\cite{Alekseev}. In a degenerate 2D electron gas at zero temperature, the inverse electron-electron scattering lifetime, $\tau_{ee}^{-1}\propto\xi_{\mathbf{p}}^2\ln{\xi_{\mathbf{p}}}$, turns into zero at the Fermi surface ($\xi_{\mathbf{p}}\rightarrow0$). In the case of the Bose-Fermi mixture, the unpaired electron lifetime, $\tau_e$, which comes from the electron-boson scattering, also makes the contribution to the viscosity. It follows from Eq.~\eqref{lifetime4} that  $\tau_e^{-1}\propto\xi_{\mathbf{p}}$ and it turns into zero more slowly than $\tau_{ee}^{-1}$ at $\xi_{\mathbf{p}}\rightarrow0$. This means that the Fermi subsystem viscosity is determined by the fermion-boson scattering processes rather than by the fermion-fermion ones. As a consequence, one can expect that the superfluid Bose subsystem will give the predominant contribution to the inter-subsystem viscosity in comparison with the fermion-fermion interaction.

\subsection{Collective modes}


In the collective modes, the fermion density fluctuations $\delta n_{\textbf{k}\omega}$ and the boson density fluctuations $\delta N_{\textbf{k}\omega}$ are coupled by the system of equations
\begin{eqnarray}\label{CollModes1}
\delta n_{\textbf{k}\omega}&=&S_{\textbf{k}\omega}U^{BF}_{\textbf{k}} \delta N_{\textbf{k}\omega},\\\nonumber
\delta N_{\textbf{k}\omega}&=&P_{\textbf{k}\omega}U^{BF}_{\textbf{k}} \delta n_{\textbf{k}\omega},
\end{eqnarray}
where $U^{BF}_{\textbf{k}}$ is the Fourier transform of the boson-fermion interaction potential $U_{BF}(\textbf{r})$,
\begin{gather}\label{CollModes2}
S_{\textbf{k}\omega}=\frac{\Pi_{\textbf{k}\omega}}{1-U^F_\mathbf{k}\Pi_{\textbf{k}\omega}},\,\,
P_{\textbf{k}\omega}=\frac{n_ck^2/M}{(\omega+i\delta)^2-\omega_k^2}
\end{gather}
are the Fermi subsystem response function and the Bose subsystem response function, respectively, which describe the reaction of the subsystems to an external perturbation,
\begin{gather}\label{CollModes3}
\Pi_{\textbf{k}\omega}=-\frac{m}{\pi}\left[1-\frac{|\omega|\theta[\omega^2-v_F^2k^2]}{\sqrt{\omega^2-v_F^2k^2}}-i\frac{|\omega|\theta[v_F^2k^2-\omega^2]}{\sqrt{v_F^2k^2-\omega^2}}\right],
\end{gather}
is the Fermi subsystem polarization operator written in the long wavelength limit ($k\ll mv_F$), and $\omega_k=sk$ is the bogolon dispersion. The poles of the response functions \eqref{CollModes2} give the dispersions of the corresponding collective modes in the system.  Namely, the $P_{\textbf{k}\omega}$ pole, $\omega=\omega_k$, defines the Bogoliubov mode, whereas the $S_{\textbf{k}\omega}$ pole defines the plasmon mode. It should be noted that the plasmon mode exists only within the frequency domain $\omega\gg kv_F$, where the imaginary part of the polarization operator \eqref{CollModes3} is $\mathrm{Im}\,\Pi_{\textbf{k}\omega}=0$ and its real part can be written as $\mathrm{Re}\,\Pi_{\textbf{k}\omega}\approx mv_F^2k^2/2\pi \omega^2$. As a result, the denominator of the response function $S_{\textbf{k}\omega}$ reads $1-U^F_\mathbf{k}Re\,\Pi_{\textbf{k}\omega}\approx1-\omega_p^2/\omega^2$ and has the pole $\omega=\omega_p$}, where $\omega_p\equiv v_F\sqrt{k_sk/2}$ is the plasmon dispersion.

The secular equation of the algebraic system~\eqref{CollModes1} yields the dispersion equation describing the interaction between the plasmon and Bogoliubov modes,
\begin{gather}\label{CollModes4}
1-U^F_\mathbf{k}\Pi_{\textbf{k}\omega}-(U^{BF}_{\textbf{k}})^2\Pi_{\textbf{k}\omega}P_{\textbf{k}\omega}=0,
\end{gather}
which can be rewritten within the domain $\omega\gg kv_F$ as
\begin{gather}\label{CollModes5}
(\omega^2-\omega_p^2)(\omega^2-\omega_k^2)-(\omega_p\omega_k)^2k_s/k=0.
\end{gather}
Solving Eq.~\eqref{CollModes5}, we arrive at the hybridized plasmon-bogolon modes,
\begin{gather}\label{CollModes6}
\omega^2_{1,2}=\frac{\omega_p^2+\omega_k^2}{2}\pm\frac{1}{2}\sqrt{(\omega_p^2-\omega_k^2)^2+4(\omega_p\omega_k)^2\frac{k_s}{k}},
\end{gather}
written in the limit $k\ll k_s$. It should be noted that the mode $\omega_2$ does not exist physically since $\mathrm{Re}\,\omega_2=0$. On the contrary, the hybridized mode $\omega_1$ is not damped since $\textmd{Im}\,\Pi_{\textbf{k}\omega}=0$ for $\omega>kv_F$ and $\textmd{Im}\,P_{\textbf{k}\omega}\propto\delta(\omega^2-\omega_k^2)=0$ for $\omega_{1}\neq\omega_k$.
\begin{figure}[t]
\includegraphics[width=0.8\columnwidth]{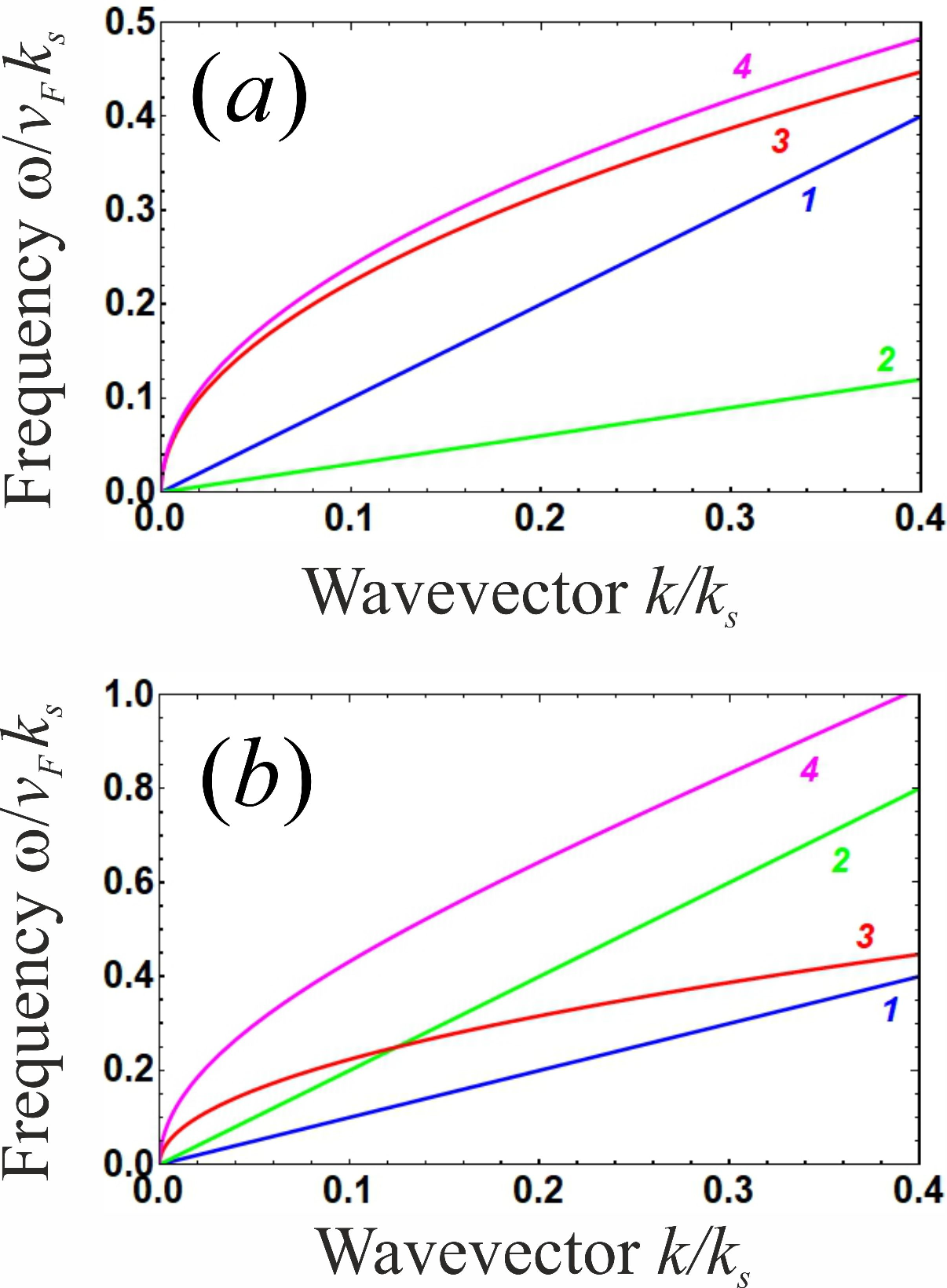}
\caption{Dispersion of the collective modes for the different values of the Bogoliubov phase velocity: (a) $s<v_F$;  (b) $s>v_F$. The line $1$ corresponds to the dispersion $\omega=v_Fk$, the line $2$ is the bare Bogoliubov mode dispersion $\omega_k=sk$, the line $3$ is the bare plasmon dispersion $\omega_p$, and the line $4$ is the hybridized plasmon-bogolon mode $\omega_1$.}
\label{Fig.7}
\end{figure}
The hybridized $\omega_1$ mode is plotted for the cases of $s<v_F$ and $s>v_F$ in Fig.~6. It follows from the plots that the hybridization of the plasmon and Bogoliubov modes is most pronounced if the Bogoliubov mode velocity exceeds the Fermi velocity, i.e. $s>v_F$ (see Fig.~6b). In this case the Bogoliubov mode (line $2$) and the plasmon mode (line $3$) are crossed and, therefore, their interaction is most effective. In the opposite case, $s<v_F$, the intermode influence is relatively weak since the Bogoliubov mode (line $2$) and the plasmon mode (line $3$) are widely separated in frequencies (see Fig.~6a). As a consequence, the ultraviolet shift of the hybridized mode $\omega_1$ (line $4$) with respect to the bare plasmon dispersion (line $3$) for $s>v_F$ (see Fig.~6b) is much larger as compared with the same shift for $s<v_F$ (see Fig.~6a).


In the frequency domain $\omega < v_Fk$, the bare plasmon does not exist since the real part of the polarization operator, $\mathrm{Re}\,\Pi_{\textbf{k}\omega}$, does not depend on frequency. Therefore, only the Bogoliubov mode survives there. However, the Bogoliubov mode experiences damping in the region below the line $\omega=kv_F$ (see the lines $2$ and $3$ in Fig.~6b). The imaginary correction to the Bogoliubov mode dispersion, which arises from  $\textmd{Im}\,\Pi_{\textbf{k}\omega}\neq0$, can be easily found from the dispersion equation~\eqref{CollModes4} in the limit of $\omega\ll v_Fk$. In this limiting case, the fermion polarization operator~\eqref{CollModes3} can be simplified as
\begin{gather}\label{damping1}
\Pi_{\textbf{k}\omega}\approx-\frac{m}{\pi}\left[1-i\frac{|\omega|}{v_Fk}\right].
\end{gather}
Then the dispersion equation
\begin{gather}\label{damping2}
1-(U^{BF}_{\textbf{k}})^2S_{\textbf{k}\omega}\frac{n_ck^2/M}{\omega^2-\omega_k^2}=0
\end{gather}
yields the imaginary correction to the frequency,
\begin{align}\label{damping3}
&\textmd{Im}\,\omega=(U^{BF}_{\textbf{k}})^2\frac{n_ck^2}{2M\omega_k}\textmd{Im}\,S_{\textbf{k},\omega=\omega_k}\nonumber\\
&=2\pi\frac{\hbar^2n_c}{mv_FMs}\omega_k,
\end{align}
which describes the Bogoliubov mode damping. Depending on the boson and fermion density values, the damping can be both strong ($\lim\limits_{k\rightarrow 0}\textmd{Im}\,\omega/\omega_k\gg1$) and weak ($\lim\limits_{k\rightarrow 0}\textmd{Im}\,\omega/\omega_k\ll1$). An estimation for the MoS$_2$ monolayer with the BEC density $n_c=4\cdot10^{10}$~cm$^{-2}$ and the fermion density $n=4\cdot10^{12}$~cm$^{-2}$ results in $\lim\limits_{k\rightarrow 0}\textmd{Im}\,\omega/\omega_k=0.035$, what corresponds to the small damping of the Bogoliubov modes.

The knowledge of the dispersion laws and the damping of collective modes is the key thing in using the Bose and Fermi systems as active elements of the plasmonics~\cite{plasmonics}. Since plasmons are accompanied by the electron gas polarization, they are extremely sensitive to external electromagnetic fields. Therefore, the discussed field-induced effects can be of interest for creating high-performance plasmonic devices and technologies. In the case of light-induced hybrid Bose-Fermi systems, both bare collective excitations (plasmons and the Bogoliubov modes) and their hybrid counterparts can be studied via the well developed pump-probe experimental technique, where the strong pump field produces the light-heavy electron pairs, whereas the relatively weak probe field may excite the hybrid modes. It should be noted that the bare plasmons in conventional systems are sensitive to the electron-impurity scattering which results in the plasmon damping and widening the plasmon resonance. One can expect that the hybridized modes considered above will be less sensitive to this destructive effect since the damping of the Bogoliubov modes due to impurity scattering is weak~\cite{KovalevChaplik}. It should be noted also that the found structure of the collective modes will be useful to describe the gauge-invariant current response of the Bose-Fermi systems in the superconducting regime~\cite{Arseev}.

\section{Conclusion}

We have developed the theory describing various physical characteristics --- including the dispersion laws and the damping (lifetimes) of both single-particle and collective elementary excitations --- in the hybrid Bose-Fermi system induced by light in the two-dimensional systems containing charge carriers with different effective masses. It is shown, particularly, that the interaction between the Bose and Fermi subsystems leads to increasing effective mass of fermions (the polaron effect), the bogolon emission by a moving fermion (the Cherenkov-like effect), and the hybridization of collective modes in the Fermi subsystem (plasmons) and the Bose subsystem (bogolons). These effects can be observed in various 2D structures containing charge carriers with different effective masses, including MoS$_2$ monolayers (where the conduction band consists of the spin-split heavy electron subbands and light electron subbands) and hole systems in quantum wells based on semiconductor materials (where the valence band consists of the heavy hole subbands and light hole subbands).

\begin{acknowledgments}
The reported study was funded by the Russian Science Foundation (project 20-12-00001).
\end{acknowledgments}

\appendix
{
\section{The two-electron Hamiltonian}

Let us consider a 2D structure containing the two electron subbands with the different effective masses $m_l$ and $m_h$ (see Fig.~2), where the energy spectrum of the subbands is $\varepsilon_{l}(\mathbf{k})=-\Delta_0/2+\hbar^2k^2/2m_{l}$ and $\varepsilon_{h}(\mathbf{k})=\Delta_0/2+\hbar^2k^2/2m_{h}$,  $\mathbf{k}=(k_x,k_y)$ is the momentum of charge carrier in the 2D plane, and $\Delta_0$ is the energy splitting of the subbands at $\mathbf{k}=0$. In the presence of a circularly polarized electromagnetic wave incident normally to the 2D structure (see Fig.~1), the Coulomb interaction of two electrons from the subbands $\varepsilon_{l}(\mathbf{k})$ and $\varepsilon_{h}(\mathbf{k})$ is described by the Hamiltonian
\begin{equation}\label{H}
\hat{\cal H}=\hat{\cal H}_{l}+\hat{\cal H}_{h}+U(\mathbf{r}_{l}-\mathbf{r}_{h}),
\end{equation}
where $\hat{\cal H}_{l,h}=(\hat{\mathbf{p}}_{l,h}-e\mathbf{A}(t)/c)^2/2m_{l,h}$ are the Hamiltonians of free electrons irradiated by the wave, $\mathbf{r}_{l,h}=(x,y)$ are the plane radius vectors of the electrons, $\hat{\mathbf{p}}_{l,h}=-i\hbar\partial/\partial\mathbf{r}_{l,h}$ are the plane momentum operators of the electrons, $U(\mathbf{r}_{l}-\mathbf{r}_{h})=e^2/\epsilon|\mathbf{r}_{l}-\mathbf{r}_{h}|$ is the two-dimensional Coulomb potential of the electron interaction, $\epsilon$ is the dielectric constant,
\begin{equation}\label{A}
\mathbf{A}(t)=(A_x,A_y)=[cE_0/\omega_0](\cos\omega_0 t,\,\sin\omega_0
t)
\end{equation}
is the vector potential of the wave, $E_{0}$ is the electric field amplitude of the
wave, and $\omega_0$ is the wave frequency. The Hamiltonian \eqref{H} is spinless since the exchange interaction of the considered two electrons is absent due to different masses of them, whereas their direct spin-spin interaction is relativistically small and can be neglected as a first approximation~\cite{Kibis_2021_3}. Taking into account Eq.~\eqref{A}, the Hamiltonian \eqref{H} can be rewritten as
\begin{eqnarray}\label{H1}
\hat{\cal H}&=&\frac{\hat{\mathbf{p}}_{l}^2}{2m_l}+\frac{\hat{\mathbf{p}}_{h}^2}{2m_h}
-\frac{e\mathbf{A}(t)\hat{\mathbf{p}}_{l}}{cm_l}-\frac{e\mathbf{A}(t)\hat{\mathbf{p}}_{h}}{cm_h}
+\varepsilon_0\nonumber\\
&+&U(\mathbf{r}_{l}-\mathbf{r}_{h}),
\end{eqnarray}
where
\begin{equation}\label{e0}
\varepsilon_0=\frac{e^2E_0^2}{2m_l\omega_0^2}+\frac{e^2E_0^2}{2m_h\omega_0^2}
\end{equation}
is the kinetic energy of electron rotation under the circularly polarized field \eqref{A}. To proceed, let us apply the Kramers-Henneberger unitary transformation,
\begin{align}\label{KH}
&\hat{U}(t)=\exp\left\{\frac{i}{\hbar}\int^{\,t}\left[
\frac{e}{m_lc}\mathbf{A}(\tau)\hat{\mathbf{p}}_l-\frac{e^2E_0^2}{2m_l\omega_0^2}\right]d\tau\right\}\nonumber\\
&\times\exp\left\{\frac{i}{\hbar}\int^{\,t}\left[
\frac{e}{m_hc}\mathbf{A}(\tau)\hat{\mathbf{p}}_h-\frac{e^2E_0^2}{2m_h\omega_0^2}\right]d\tau\right\}.
\end{align}
Then the transformed Hamiltonian \eqref{H1} reads
\begin{eqnarray}\label{H11}
\hat{\cal H}^{\prime}&=&\hat{U}^\dagger(t)\hat{\cal H}\hat{U}(t) -
i\hbar\hat{U}^\dagger(t)\partial_t
\hat{U}(t)\nonumber\\
&=&\frac{\hat{\mathbf{p}}^2_l}{2m_l}+\frac{\hat{\mathbf{p}}^2_h}{2m_h}+U\big(\mathbf{r}_l-\mathbf{r}_h-
\mathbf{r}_0(t)\big),
\end{eqnarray}
where
\begin{equation}\label{r0}
\mathbf{r}_0(t)=(-r_0\sin\omega_0 t,\,r_0\cos\omega_0 t)
\end{equation}
is the vector defining the change of relative position of the two electrons under the field, and
\begin{equation}\label{r01}
r_0=\frac{|e|E_0(m_h-m_l)}{m_lm_h\omega_0^2}
\end{equation}
is the length of the vector. It should be noted that the field-induced energy \eqref{e0} results only in the energy shift of all electronic states by the same energy. Since such a shift does not affect electronic properties, the unitary transformation \eqref{KH} removes the energy \eqref{e0} from the Hamiltonian \eqref{H11}. In the center-of-mass system, the two-electron Hamiltonian \eqref{H11} can be rewritten as
\begin{equation}\label{Hcm}
\hat{\cal H}^{\prime}=\frac{\hat{\mathbf{p}}\,^2}{2m^\ast}+U\big(\mathbf{r}-
\mathbf{r}_0(t)\big),
\end{equation}
where $\mathbf{r}=\mathbf{r}_l-\mathbf{r}_h$ is the radius vector describing the relative motion of electrons, $\hat{\mathbf{p}}=-i\hbar\partial/\partial\mathbf{r}$ is the momentum operator corresponding to the relative motion, and $m^\ast=m_lm_h/(m_l+m_h)$ is the reduced mass of the two-electron system.

It should be noted that the Hamiltonian \eqref{Hcm} with the periodically time-dependent potential $U\big(\mathbf{r}-
\mathbf{r}_0(t)\big)$ is still exact and describes the relative motion of two interacting electrons under the field accurately. Next, let us apply the high-frequency approximation which is well-known in the Floquet theory of periodically driven quantum systems~\cite{Goldman_2014,Bukov_2015,Eckardt_2015,Casas_2001}. Namely, the periodically time-dependent potential in the Hamiltonian  \eqref{Hcm} can be replaced approximately with the time-averaged potential if the field frequency is high enough~\cite{Kibis_2019,Kibis_2020_2,Kibis_2021_3}. Within this approximation, the periodically time-dependent Hamiltonian \eqref{Hcm} turns into the effective stationary Hamiltonian
\begin{equation}\label{H0}
\hat{\cal H}_0=\frac{\hat{\mathbf{p}}\,^2}{2m^\ast}+U_0(\mathbf{r}),
\end{equation}
where the time-averaged potential
\begin{align}\label{U0}
&U_0(\mathbf{r})=\frac{1}{2\pi}\int_{-\pi}^{\pi}U\big(\mathbf{r}-\mathbf{r}_0(t)\big)\,d(\omega_0
t)\nonumber\\
&=\left\{\begin{array}{rl}
({2e^2}/{\pi r_0})K\left({r}/{r_0}\right),
&{r}/{r_0}\leq1\\\\
({2e^2}/{\pi r})K\left({r_0}/{r}\right),
&{r}/{r_0}\geq1
\end{array}\right.
\end{align}
can be treated as the Coulomb potential dressed by the circularly polarized field, and the function $K(z)$ is the complete elliptical integral of the first kind. Since the dressed potential \eqref{U0} has a local minimum at ${r}=0$ for $m_l\neq m_h$, the Schr\"odinger equation with the Hamiltonian \eqref{H0} yields the bound two-electron state localized near the minimum (composite boson)~\cite{Kibis_2019}. It should be noted that this bound state is quasi-stationary since the potential minimum at ${r}=0$ is local. Therefore, a single composite boson has finite lifetime. However, it has been demonstrated that the Fermi sea of normal electrons stabilizes the boson~\cite{Kibis_2021_3}. Such a stabilization is physically similar to the stabilization of the Cooper pair by the Fermi sea of conduction electrons in the conventional BCS theory of superconductivity. As a consequence, the light-induced composite bosons in the hybrid Bose-Fermi system have infinite lifetime and the system as a whole is stable~\cite{Kibis_2021_3}.

It follows from Eq.~\eqref{U0} that the dressed potential $U_0(\mathbf{r})$ for $m_l=m_h$ turns into the bare Coulomb potential, $U(\mathbf{r})=e^2/\epsilon{r}$, which has no local minima and, therefore, cannot couple interacting electrons. Physically, this follows from the fact that the vector \eqref{r0} turns into zero if $m_l=m_h$. As a consequence, the field does no change the distance between interacting electrons in this case and, correspondingly, does not affect the Coulomb interaction of them. Therefore, the condition $m_l\neq m_h$ is crucial for the effects under consideration. Among 2D structures satisfying this condition, both MoS$_2$ monolayers (where the conduction band consists of the spin-split heavy electron subbands and light electron subbands) and hole systems in quantum wells based on semiconductor materials (where the valence band consists of the heavy hole subbands and light hole subbands) should be noted.

Next, let us discuss interactions in the light-induced hybrid Bose-Fermi system, assuming the boson density to be small enough to consider the composite bosons as weakly interacting independent particles. Since the dressed Coulomb potential \eqref{U0} turns into the bare Coulomb potential for charge particles with identical masses, the boson-boson interaction and the fermion-fermion interaction can be described by the bare Coulomb potentials \eqref{FF} and \eqref{BB}, respectively. Since the boson and fermion masses are different, the boson-fermion interaction, rigorously, should be described by the dressed Coulomb potential. However, the dressed potential \eqref{U0} substantially differs from the bare Coulomb potential only for small distances $r\alt r_0$, where the length $r_0$ defined by Eq.~\eqref{r01} is the characteristic size of composite boson~\cite{Kibis_2021_3}. Therefore, the boson-fermion interaction for small boson densities can be described by the bare Coulomb potential defined by Eq.~\eqref{BF}.}

\end{document}